\documentclass[useAMS,usenatbib]{mn2e}
\usepackage{graphicx}
\usepackage{multirow}
\usepackage{txfonts}

\title[The hybrid B-type pulsator $\gamma$ Pegasi]{The hybrid B-type pulsator $\gamma$ Pegasi: mode identification and complex seismic modelling}
\author[P. Walczak, J. Daszy\'nska-Daszkiewicz, A.A. Pamyatnykh, T. Zdravkov]{P. Walczak$^{1}$\thanks{E-mail: walczak@astro.uni.wroc.pl}, J. Daszy\'nska-Daszkiewicz$^{1}$\thanks{E-mail: daszynska@astro.uni.wroc.pl}, A.A. Pamyatnykh$^{2}$\thanks{E-mail: alosza@camk.edu.pl},  T. Zdravkov$^{2}$\thanks{E-mail: songo@camk.edu.pl} \\
$^{1}$Instytut Astronomiczny, Uniwersytet Wroc{\l}awski, ul. Kopernika 11, 51-622 Wroc{\l}aw, Poland\\
$^{2}$Nicolaus Copernicus Astronomical Center, Bartycka 18, 00-716, Warsaw, Poland\\
}
\begin{document}

\date{Accepted ... Received ...; in original form ...}

\pagerange{\pageref{firstpage}--\pageref{lastpage}} \pubyear{2002}

\maketitle

\label{firstpage}

\begin{abstract}
We present interpretation of the oscillation spectrum of the early B-type star $\gamma$ Pegasi, in which both low order p/g and high-order g-modes are observed.
Using amplitudes and phases of the photometric and radial velocity variations, we identify/constrain the mode degree, $\ell$,
for all 14 detected frequencies. Seismic models fitting two pulsational frequencies corresponding to the modes $\ell=0$, p$_1$ and $\ell=1$, g$_1$
were constructed. This set of models contains those which reproduce also the empirical values of the complex nonadiabatic parameter $f$ associated
to these two mode frequencies. Unfortunately, there are no models reproducing the values of $f$ for both frequencies simultaneously, regardless of model
atmospheres, opacity data, chemical mixture as well as opacity enhancement in the $Z-$ and Deep Opacity Bumps.
 Most probably, some modifications of the opacities in stellar interiors are still required.
\end{abstract}


\section{Introduction}

The simultaneous presence of both acoustic (p) and internal gravity (g) modes in a star,
allows potentially for probing almost the entire interior.
Such hybrid oscillations have been detected in a few main sequence stars of early B spectral type: $\nu$ Eridani \citep{Hetal04, Aetal04, J05}, 12 Lacertae \citep{Hetal06,Detal09}, $\gamma$ Pegasi \citep{Hetal09}, HD50230 \citep{Degroote_al2012}, HD43317 \citep{Papics_al2012}
and many candidates have been found (Pigulski\& Pojma\'nski 2008, Degroote et al. 2009, Balona et al. 2011).
It was only possible thanks to multisite campaigns and space observations from the satellite missions like MOST \citep{MOST}, CoRoT \citep{Corot} and Kepler \citep{Kepler}.
The existence of both low order p/g modes as well as high-order g-modes makes the early B-type pulsators the attractive
targets for asteroseismic studies.


The basic goal of asteroseismology is to find pulsational models with eigenfrequency reproducing the observational values.
A more advanced seismic model should additionally account for instability and properties of oscillation modes.
\citet{DDDP03,DDDP05} introduced a new asteroseismic tool associated with each pulsational frequency.
It is a ratio of the amplitude of the bolometric flux perturbations to the radial displacement, called the nonadiabatic $f$-parameter.
The theoretical values of $f$ are obtained from stellar pulsation computations and their empirical
counterparts are derived from multicolour photometry and radial velocity data simultaneously with the mode degree, $\ell$.
The parallel fitting of pulsational frequencies and corresponding values of the $f$-parameter was termed {\it complex asteroseimology} by \citet{DW09}.
The hybrid pulsators are of particular interest for such an in-depth modelling because
the $f$-parameter depends on pulsational frequency and the shape of eigenfunctions. Therefore its
behavior is very different for the different frequency ranges.
In the case of low order p/g modes (the high frequency range), the values of $f$ are independent of the mode degree $\ell$
and slowly vary with the frequency. On the contrary, for high order g-modes (the low frequency range), the values of $f$
are strongly $\ell$-dependent and change rapidly with the frequency.

Complex seismic modeling has been already applied to the $\beta$ Cep star $\theta$ Oph \citep{DW09}
and to the two $\beta$ Cep/SPB stars: $\nu$ Eri \citep{DW10} and 12 Lac (Daszy\'nska-Daszkiewicz, Szewczuk \& Walczak 2013).
In the case of $\theta$ Oph, we got a strong preference for the OPAL \citep{OPAL} opacity tables.
From the analysis of $\nu$ Eri a contradictory result was obtained: the $\beta$ Cep-type modes indicated the OPAL data
whereas the SPB-type modes preferred the OP \citep{OP} opacities. For 12 Lac, somewhat better agreement was found with the OP data. It can indicate, that the used opacity tables are not properly defined in the whole range of temperature and various physical conditions. Up to now, the sample of analyzed stars is too small to draw more general conclusions.

In this paper, we present mode identification and complex seismic modeling of $\gamma$ Pegasi. The analysis of the space based observations from the MOST satellite \citep{Hetal09} and ground based photometry and spectroscopy \citep{H09}
led to the discovery of 14 pulsational frequencies with 8 typical for the $\beta$ Cephei stars and 6 typical for the Slowly Pulsating B-type stars (SPB). Because the star is a very slow rotator, here, we neglect all effects of rotation on pulsations.

In Section 2, we give a short description of the star. Section 3 contains results on
mode identification for all detected pulsational frequencies using two approaches.
Results of our seismic modeling are presented in Section 4.
Conclusions are summarised in Section 5.

\section{The hybrid pulsator: $\gamma$ Pegasi}

$\gamma$ Peg (HR 39, HD 886) is a pulsating star of B2IV spectral type with the brightness of V=2.83 mag. The radial velocity variations were discovered by \cite{B1911}
but a very small amplitude made the variability uncertain. Series of spectrograms obtained by \cite{McN53} showed, that this star is a variable with the period
of about $3^{\rm{h}}38^{\rm{m}}$ and the amplitude of the radial velocity variations of 3.5 km/s. \citet{McN53} attributed $\gamma$ Peg to the $\beta$ Cephei class
of variable stars on the basis of its short period and spectral type. The follow-up spectroscopic and photometric observations confirmed changes of the radial
velocity \citep{McN1955,McN1956,SMcN1960} and led to the discovery of the light variations in the yellow filter with the range of 0.015 mag \citep{W1954}.
\citet{J1970} determined the following light ranges in the UBV passbands: $0.027\pm0.001$, $0.018\pm0.001$ and $0.017\pm0.002$ mag, respectively.
\citet{Setal1975} determined period more precisely, $P=0.1517501(3)$ d, and confirmed its constancy within 0.06 s per century.
\cite{SMc1978} and \cite{Cetal1994} showed, that the variability of $\gamma$ Peg can be explained by the radial pulsation.

In the catalogue of the galactic $\beta$ Cephei stars by \citet{SH05}, $\gamma$ Peg is the lowest mass variable,
close to the Slowly Pulsating B stars (SPB) instability strip. $\gamma$ Peg had been thought to be one of the monoperiodic
$\beta$ Cep star until \citet{C06} reported three additional pulsational frequencies; one in the $\beta$ Cephei range and two
in the SPB frequency domain. \citet{C06} claimed also that the star is a spectroscopic binary with an orbital period of 370.5 d.
However, the analysis of data collected by the MOST satellite and from ground-based photometric and spectroscopic observations
\citep{Hetal09,H09}, showed that $\gamma$ Peg is a single star and the hypothetical orbital variations can be explained by the
high-order g-mode pulsation. Moreover, \citet{Hetal09} confirmed the frequencies of \citet{C06} and discovered ten new ones.
Consequently, we know that the light variations of $\gamma$ Peg are caused by at least 14 pulsational frequencies: 8 of them are of the $\beta$ Cep type
and 6 are of the SPB type. Thus, $\gamma$ Peg is the hybrid pulsator in which low-order pressure/mixed modes
 and high-order gravity modes are excited simultaneously.

The rotational velocity of $\gamma$ Peg was determined by \citet{Tetal06} who derived $V_{\rm{rot}}\sin{i}\approx 0$
from the SiIII lines. Similar result was given by Pandey et al. (2011) and \citet{Nieva+2012}, who got $V_{\rm rot}=6$ km/s and $V_{\rm rot}=9\pm 2$ km/s, respectively. \citet{Hetal09} assuming that two pulsational frequencies are components of the $\ell=1$, g$_1$ triplet,
determined $V_{\rm{rot}}\approx 3$ km/s from the rotational splitting.

Determinations of the heavy elements abundance are ambiguous. \citet{DDN05} obtained the metal abundance of [m/H]$=-0.04\pm0.08$
(equ\-ivalent to the metallicity parameter of $Z\approx 0.017\pm 0.004$) from the IUE ultraviolet spectra. From the optical spectra, \citet{Metal06} and \citet{Pandey_al2011}
derived $Z=0.009\pm0.002$ and $Z=0.010\pm0.002$, respectively. The recent determinations from the optical and ultraviolet IUE spectra give
$Z=0.017\pm0.001$ \citep{Wu+2011} and $Z=0.014\pm0.002$ \citep{Nieva+2012}, while the analysis of the ultraviolet HST spectra
indicates $Z=0.016\pm0.003$ \citep{Koleva+2012}.

\begin{figure}
\includegraphics[clip,width=83mm]{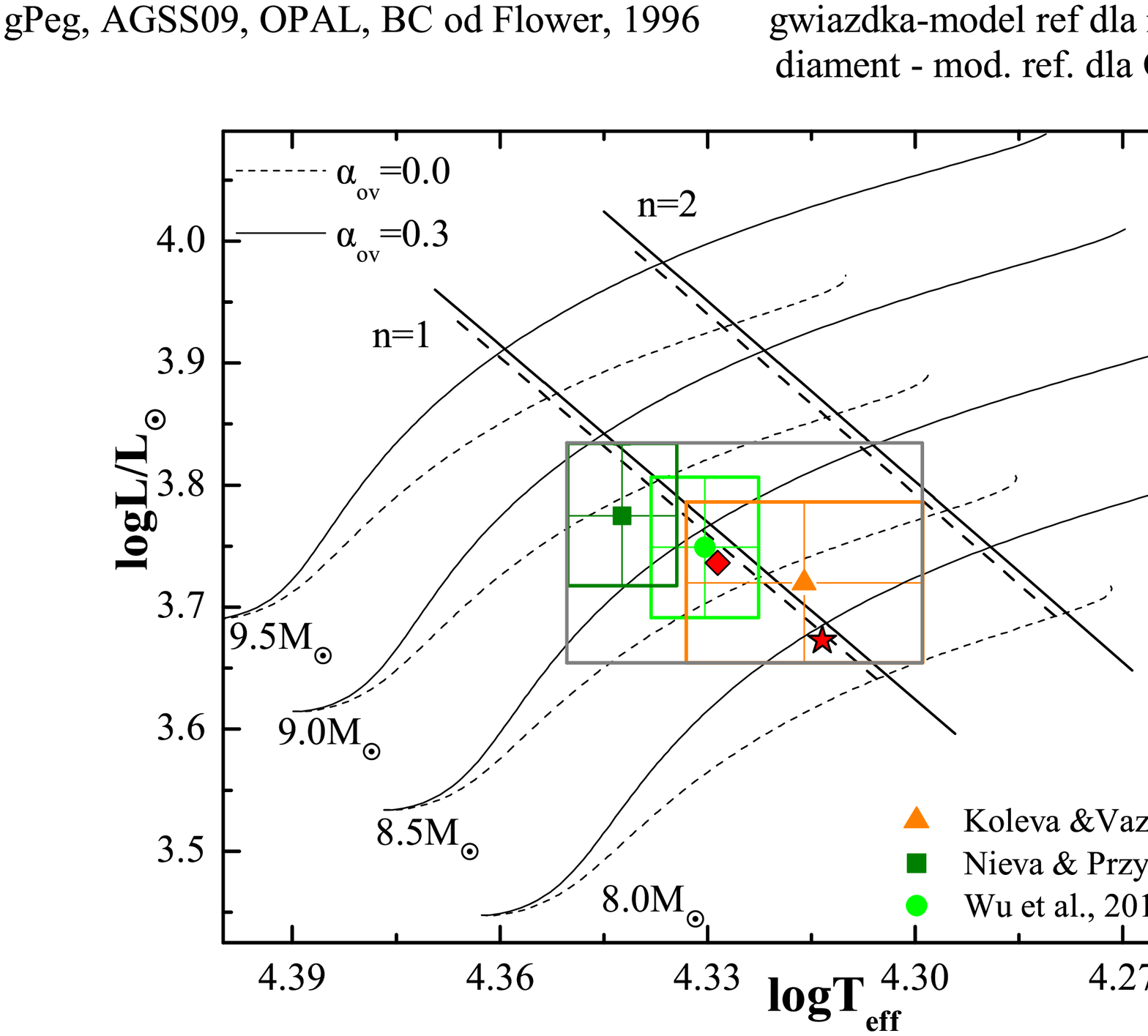}
\caption{The observational error boxes of $\gamma$ Peg in the HR diagram. The evolutionary tracks were computed
for metallicity of $Z=0.015$ and two values of the overshooting parameter, $\alpha_{\rm{ov}}$. Lines of the constant
period (0.15175 d) for the fundamental ($n=1$) and first overtone ($n=2$) radial mode are also drawn. Points labeled as a diamond and asterisk are discussed in Section\,4.}
\label{HR}
\end{figure}

In Fig.\,\ref{HR} we show the observational error box of $\gamma$ Peg in the HR diagram.  We included the most recent determination
of effective temperature by \citet{Wu+2011}, \citet{Koleva+2012} and \citet{Nieva+2012}. The luminosity was calculated from
the Hipparcos parallax, $\pi=8.33\pm0.53$ mas \citep{vanLeeuwen2007}, and the bolometric correction from Flower (1996).
The total error box is as follows: $\log{T_{\rm{eff}}}=4.325\pm0.026$ and $\log{L/L_{\odot}}=3.744\pm0.090$.

We depicted also the evolutionary tracks from the zero-age main sequence (ZAMS) to the terminal-age main sequence (TAMS) for masses
$M=8.0$, 8.5, 9.0 and 9.5$M_{\odot}$, the hydrogen abundance of $X=0.7$, metallicity of $Z=0.015$, two values of the overshooting parameter, $\alpha_{\rm{ov}} = 0.0$ and $0.3$, the initial equatorial rotational velocity of $V_{\rm{rot}}=3$ km/s and the element mixture
by Asplund et al.\,(2009), hereafter AGSS09. The tracks were computed by means of the War\-saw-New Jersey evolutionary code
\citep[e.g][]{Petal98} adop\-ting the OPAL opa\-cities \citep{OPAL}. Lines labeled as $n=1$ and $n=2$ and models marked
with a diamond and asterisk will be discussed later on.

\begin{table*}
\begin{center}
\caption{Pulsational observables of $\gamma$ Peg. In the first column, we give the pulsational frequencies of $\gamma$ Peg. In the following columns there are values of the Str\"omgren $u$, $v$, $y$ amplitudes and phases. In the last two columns there is the radial velocity amplitude and corresponding phase.}
\begin{tabular}{ccccccccccccc}
\hline
frequency &$A_u$  &$\varphi_u$ & $A_v$  &$\varphi_v$&$A_y$ &$\varphi_y$ & $A_{Vrad}$& $\varphi_{Vrad}$\\
$[$c/d$]$ & [mmag] &[rad]& [mmag]& [rad]& [mmag]&[rad]&[km/s]&[rad]\\
\hline
$\nu_{1}$=6.58974(2)&12.26(13)&2.669(11)&6.74(10)&2.628(15)&6.074(94) &2.607(15)&3.3582(69)&4.490(2) \\
$\nu_{2}$=0.63551(10)&2.44(13)&0.870(57)&1.55(10)&0.734(67)&1.269(92) &0.947(76)& 0.4992(70)&4.735(14)  \\
$\nu_{3}$=0.68241(7) &2.21(13)&5.607(61)&1.64(10)&5.681(63)&1.500(94) &5.644(63)& 0.7364(72)&3.402(10)\\
$\nu_{4}$=0.73940(10)&1.77(13)&3.542(75)&1.18(10)&3.604(89)&1.166(95) &3.782(83)& 0.5214(71)&1.468(14) \\
$\nu_{5}$=6.01616(14)&1.58(14)&4.592(87)&1.12(10)&4.664(92)&0.828(97) &4.54(12)& 0.3612(85)&0.148(24) \\
$\nu_{6}$=0.88550(7) &1.54(14)&3.733(87)&1.10(10)&4.014(94)&0.896(94) &4.00(11)& 0.7236(68)&0.873(10) \\
$\nu_{7}$=6.9776(5)  &0.48(13)&3.08(28) &0.25(10)&3.27(42)&0.328(95) &3.36(29)& 0.0962(72)&4.930(75) \\
$\nu_{8}$=0.91442(11)&0.94(13)&4.45(15) &0.60(10)&4.80(17) &0.417(95) &4.95(23)& 0.4652(68)&1.191(16) \\
$\nu_{9}$=6.5150(8)  &0.18(13)&3.46(73) &0.15(10)&2.44(68) &0.258(94) &2.64(37)& 0.0628(70)&4.15(11) \\
$\nu_{10}$=8.1861(8) &0.26(13)&4.92(50) &0.26(10)&4.89(40) &0.143(94) &5.57(66)& 0.0639(69)&1.60(11)\\
$\nu_{11}$=0.8352(3) &0.82(13)&1.30(16) &0.60(10)&0.97(17) &0.700(93) &1.43(14)& 0.1785(67)&4.743(41) \\
$\nu_{12}$=6.0273(5) &0.41(14)&1.55(34) &0.35(10)&0.53(30) &0.389(95) &0.57(25)& 0.1104(82)&1.823(75) \\
$\nu_{13}$=9.1092(12)&0.38(13)&3.61(34) &0.26(10)&3.38(39) &0.279(94) &3.24(34)& 0.0408(68)&4.51(17) \\
$\nu_{14}$=8.552(2)  &0.20(13)&0.06(67) &0.37(10)&0.59(28) &0.222(94) &5.54(42)& 0.0260(69)&1.70(27) \\
\hline
\end{tabular}
\label{Ampl}
\end{center}
\end{table*}

\section{Identification of oscillation modes}
\label{Ident}

To identify the mode degree, $\ell$, we made use of the light variations in the  Str\"omgren $uvy$ passbands \citep{H09}
and the radial velocity changes \citep{Hetal09}. In Table\,\ref{Ampl}, we give amplitudes and phases of the light and radial velocity variations determined by fitting 14 frequencies detected in the MOST data. Therefore, our values of the amplitudes and phases can differe from those determined by Handler\,(2009).

We performed identification of the degree, $\ell$, by applying two methods. In the first case, we compared theoretical and observational values of the amplitude ratios and phase differences between
the available passbands and relied on the theoretical values of the $f$-parameter. In the second approach, we used
amplitudes and phases themselves and the $f$-parameter was determined from the observations together with
the mode degree, $\ell$ \citep{DDDP03,DDDP05}. In the case of B-type pulsators, the latter method demands the radial velocity
measurements to get a unique identification of $\ell$.

In both cases we need input from model atmospheres. Here, we present results obtained with the \citet{K04} models and the microturbulent
velocity of $\xi_{t}=2$ km/s. Coefficients of the non-linear limb darkening law were adopted from \citet{Claret2000}.
Identification of $\ell$ does not change if other value of $\xi_{t}$ or the non-LTE model atmospheres \citep{Lanz2007} were used.
The theoretical values of the $f$-parameter are calculated with the nonadiabatic pulsational code of \citet{D77}.

In Table\,\ref{MI}, we give the most probable values of the mode degrees for the $\gamma$ Peg frequencies from the two approa\-ches.
The dominant mode, $\nu_1$, is certainly radial. Also the identification of the $\nu_3$, $\nu_5$ and $\nu_6$ is unique; they are
modes with $\ell=1$, 1 and 2, respectively. The frequencies $\nu_2$, $\nu_4$ and $\nu_{11}$  can be either dipoles ($\ell=1$)
or quadruples ($\ell=2$). We got also two possibilities for the frequencies $\nu_8$ and $\nu_{12}$. They can be $\ell=3$ or 2
and $\ell=2$ or 5 modes, respectively. Identification of the remaining frequencies are ambiguous. In the case of $\nu_7$, $\nu_9$,
$\nu_{10}$ and $\nu_{13}$ we can say only that their mode degrees should be different from 4 and 6,
and in the case of $\nu_{14}$ only degrees higher than 3 are possible. It is interesting that $\nu_{14}$ is close to the first
overtone radial mode in many $\gamma$ Peg models, but the degree $\ell=0$ was excluded by our photometric identification.

\begin{table}
\begin{center}
\caption{Identification of $\ell$ from the two methods. In the first column, we give the pulsational frequencies of $\gamma$ Peg,
the second and third columns contain the mode degrees, $\ell$, identified
from photometric observables using the theoretical values of $f$ and from photometric observables and radial velocity data using the empirical values of $f$, respectively. In the last column, we put values of $\ell$ consistent for both approaches.}
\begin{tabular}{lccccc}
\hline
frequency & phot.        & phot.+$V_{\rm{rad}}$ & \multirow{2}{*}{ $\ell^{*}$} \\
$[$c/d$]$ &theoret. $f$ &    empir. $f$    & \\
\hline
$\nu_{1}$=6.58974 & $\ell$=0     & $\ell$=0   & 0 \\
$\nu_{2}$=0.63551 & $\ell$=1,2,4 & $\ell$=2,1,3,5   & 2,1 \\
$\nu_{3}$=0.68241 & $\ell$=1     & $\ell$=1 & 1\\
$\nu_{4}$=0.73940 & $\ell$=1,2   & $\ell$=1,2 & 1,2 \\
$\nu_{5}$=6.01616 & $\ell$=1   & $\ell$=1,3 & 1\\
$\nu_{6}$=0.88550 & $\ell$=2     & $\ell$=2 & 2\\
$\nu_{7}$=6.9776  & $\ell$=?     & $\ell\ne 4,6$ & $\ne 4,6$\\
$\nu_{8}$=0.91442 & $\ell\le$4 & $\ell$=3,2,5 & 3,2\\
$\nu_{9}$=6.5150  & $\ell$=?   & $\ell\ne 4,6$ & $\ne 4,6$\\
$\nu_{10}$=8.1861 & $\ell$=?   & $\ell\ne 4,6$ & $\ne 4,6$\\
$\nu_{11}$=0.8352 & $\ell$=1,2   & $\ell$=1,2  & 1,2\\
$\nu_{12}$=6.0273 & $\ell$=?   & $\ell$=2,5  & 2,5\\
$\nu_{13}$=9.1092 & $\ell$=?   & $\ell\ne 4,6$ & $\ne 4,6$\\
$\nu_{14}$=8.552  & $\ell\ge 4$ &$\ell=?$ & $\ge 4$\\
\hline
\end{tabular}
\label{MI}
\end{center}
\end{table}

\begin{figure}
\begin{center}
\includegraphics[clip,width=83mm,height=63mm]{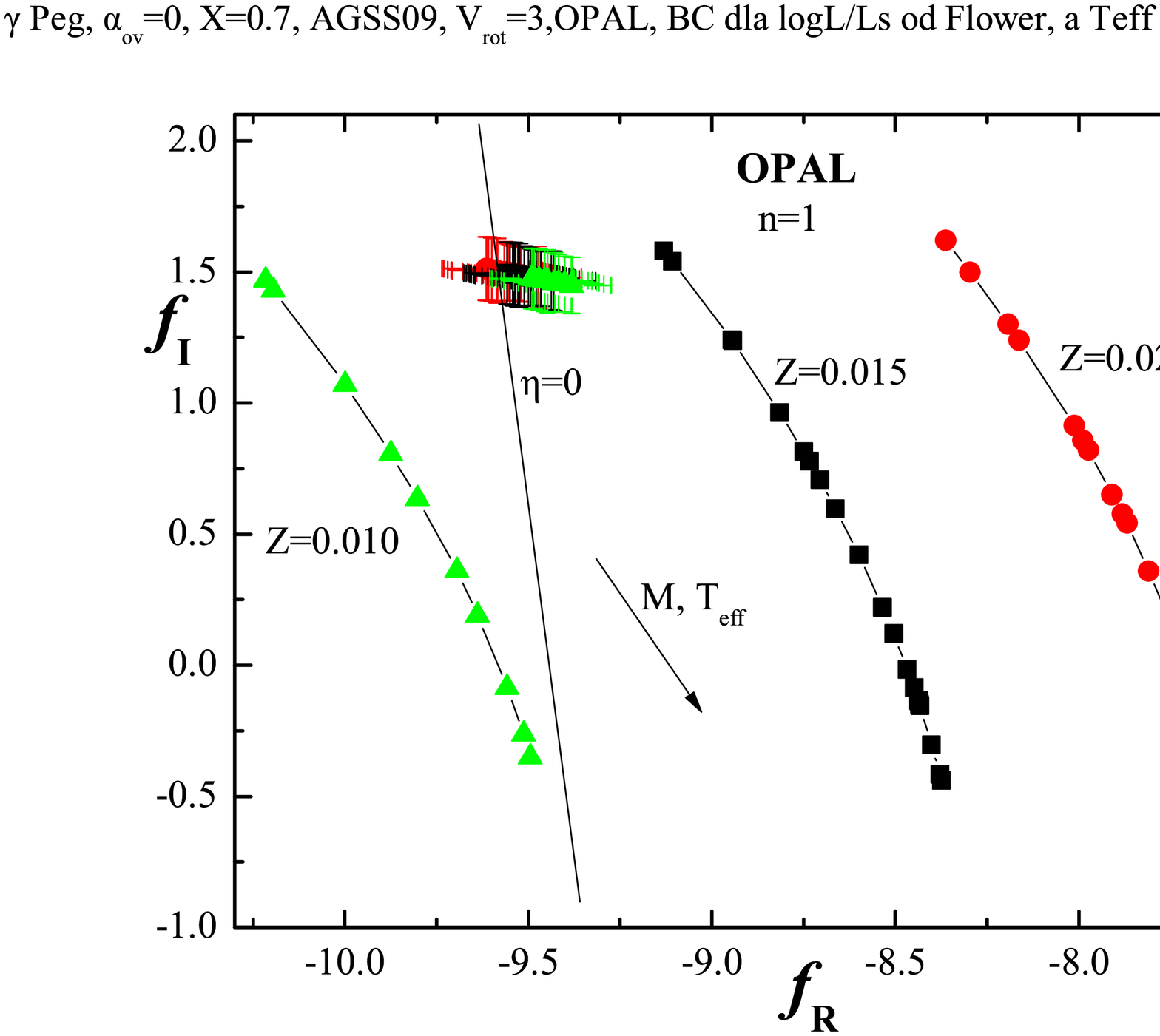}
\includegraphics[clip,width=83mm,height=63mm]{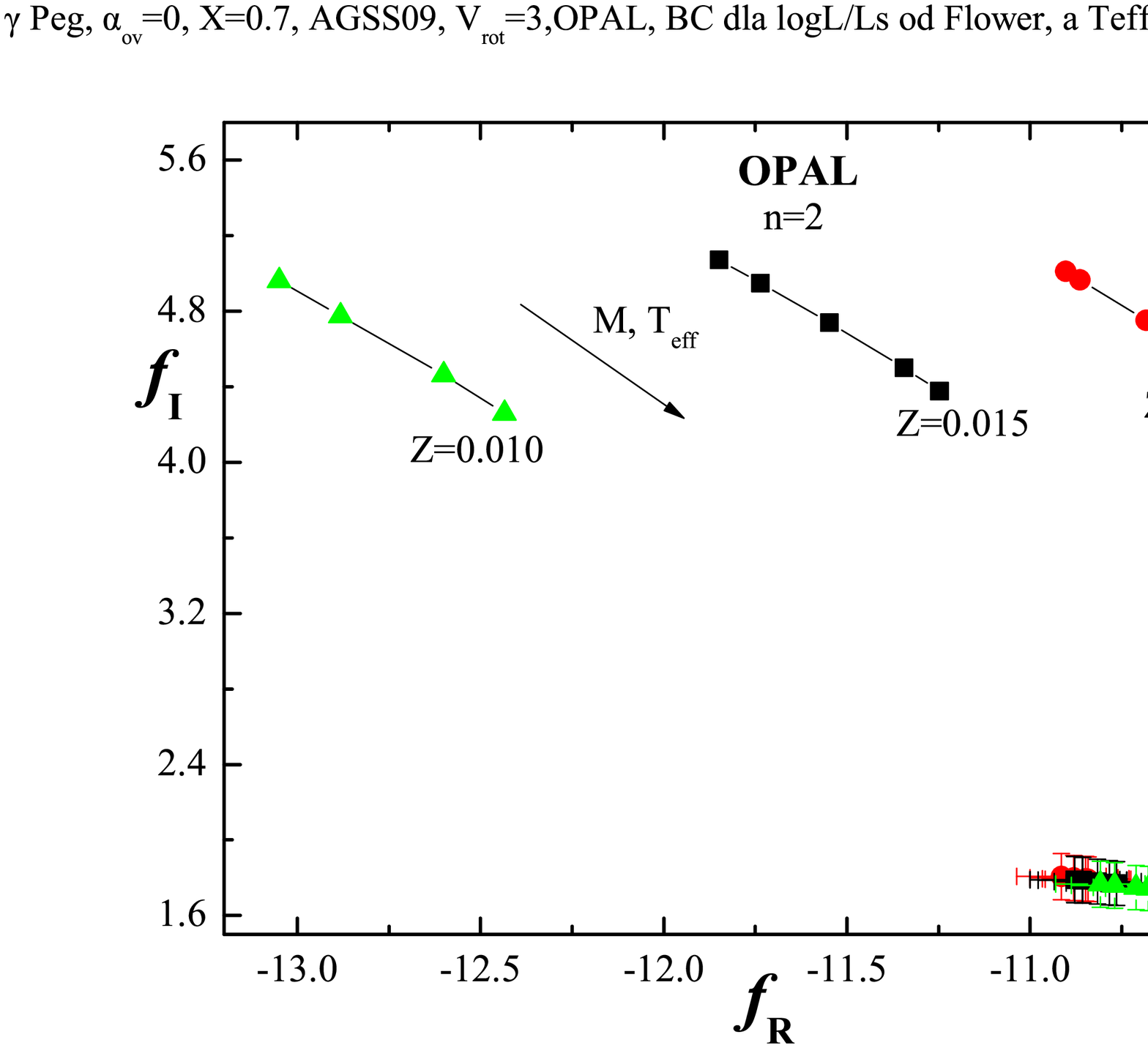}
\caption{Comparison of empirical and theoretical values of $f$ for the dominant frequency, $\nu_1$, on the complex plane ($f_{\rm{R}}$, $f_{\rm{I}}$) for models from inside the total error box. The OPAL data and AGSS09 mixture were adopted. The top and bottom panels correspond to the hypothesis of the fundamental and first overtone radial mode,
respectively. In the upper panel, models to the right of the $\eta=0$ line are stable; in the lower panel all models are stable.}
\label{fr-fi}
\end{center}
\end{figure}

In the HR diagram (Fig.\,\ref{HR}), we plot lines of a constant period (0.15175 d) corresponding to $\nu_1$ for the fundamental
and first overtone radial modes. As we can see the $p_2$ line is included only marginally in the error box. Nevertheless,
two possibilities have to be considered. The discrimination of the radial order, $n$, can be done by a comparison of the empirical
and theoretical values of the $f$-parameter (Daszy\'nska-Daszkiewicz, Dziembowski \& Pamyatnykh 2005). In Fig.\,\ref{fr-fi},
we plotted this comparison on the complex plane assuming that the dominant frequency is the fundamental (the top panel) or the first overtone
mode (the bottom panel). The theoretical values of $f$ were calculated for three different
metallicities, $Z=0.010$, 0.015, and 0.020, overshooting parameter $\alpha_{\rm{ov}}=0.0$, hydrogen abundance of $X=0.7$
and the OPAL data. In all cases, we included models that are inside the observational error box of $\gamma$ Peg.
In the upper panel, we showed the line of the constant instabillity parameter $\eta$, defined as (Stellingwerf 1978):
\begin{equation}
\eta=\frac{W}{\int_0^R \left|\frac{dW}{dr}\right|dr},
\end{equation}
where $W$ is the work integral and $R$ is the stellar radius. Models located to the right of the $\eta=0$ line are unstable, whereas in the lower panel all models are stable.
As we can see, agreement between the empirical and theoretical values of the $f$-parameter can be achieved only if $\nu_1$ is the radial fundamental mode, $n=1$.
Moreover, constraints on the metal abundance were obtained, viz $Z\in(0.012$, 0.014). The conclusion about the radial order of $\nu_1$ did not change if the OP opacities were used. In this case, we got the metallicity in the range of $Z\in(0.010, 0.012)$. Moreover, identification of the radial order is independent of the value of the core overshooting parameter, $\alpha_{\rm{ov}}$. For the higher values of $\alpha_{\rm{ov}}$, the allowed metallicity range is shifted to the smaller value of $Z$.

\section{Complex asteroseismology}

Basic seismic modeling consists in fitting pulsational frequency taking into account instability condition. This approach can be extended by a requirement of reproducing also the empirical values of the nonadiabatic parameter $f$, which corresponds to each mode frequency.

In all computations, we used the Warsaw-New Jersey stellar evolution code \citep[e.g.][]{Petal98} and nonadiabatic pulsational code
of \citet{D77}. Seismic models were calculated with both the OPAL and OP data. We assumed the initial rotational velocity of 3 km/s
and two chemical mixtures: AGSS09 and that determined for $\gamma $ Peg \citep{Nieva+2012}, hereafter the $\gamma$ Peg mixture.
Due to the excitation problem of some modes, we tested also the opacity enhancement near the $Z$-bump and Deep Opacity Bump (DOB).

\subsection{Fitting the centroid frequencies of $\gamma$ Peg}

\begin{figure}
 \begin{center}
 \includegraphics[clip,width=83mm,height=70mm]{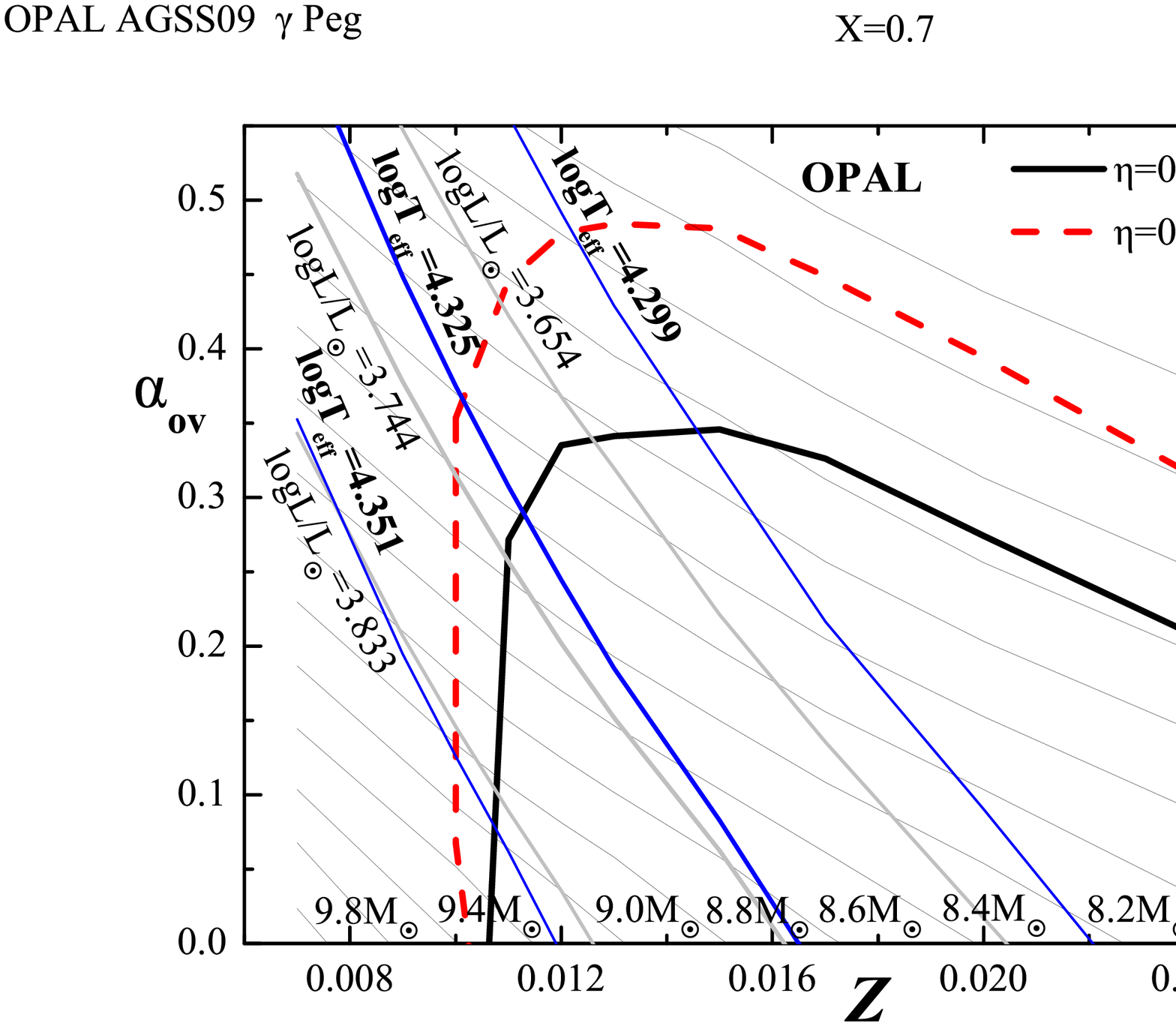}
 \caption{The overshooting parameter, $\alpha_{\rm ov}$, as a function of metallicity, $Z$, for seismic models of $\gamma$ Peg found from the fitting of the $\nu_{1}$ frequency (the $\ell = 0$, p$_1$ mode) and the $\nu_5$ frequency (the $\ell = 1$, g$_1$ mode) for the hydrogen abundance $X=0.7$ and OPAL opacities. We plot also the lines of constant mass, effective temperature and luminosity. Indicated vales of $\log{T_{\rm{eff}}}$ and $\log{L/L_{\odot}}$ are for the center and 1$\sigma$ error of the total error box. Models below the $\eta=0$ lines are unstable.}
\label{Z-AO-OPAL}
\end{center}
\end{figure}

In the first step we searched for models fitting the value of the dominant frequency, $\nu_1$, identified as the radial fundamental
mode in previous section. Next, from this set of models we selected only those, which fit also the observational value of the
frequency $\nu_5$. In the whole set of models, we found that if $\nu_1$ is the radial fundamental mode, then $\nu_5$ can be only the dipole g$_1$ mode. This identification of $\nu_5$ is unique, because the difference between frequencies of consecutive dipole modes is of the order of 1 c/d. Identification of the radial order is also independent of the azimuthal number, $m$, since the rotational splitting is of the order of 0.005 c/d. The frequencies $\nu_{1}$ and $\nu_5$ were chosen because they are well
identified low-order pressure/mixed modes. High-order gravity modes have a very dense frequency spectrum and fixing the radial order
is problematic. The accuracy of the fitting $\nu_1$ and $\nu_5$ is of the order of $10^{-5}$ c/d, which is equivalent to the observational errors.

These seismic models of $\gamma$ Peg, calculated with the hydrogen abundance of $X=0.7$, the AGSS09 chemical composition and OPAL
opacities, are shown in Fig.\,\ref{Z-AO-OPAL} on the $\alpha_{\rm{ov}}~ vs.~ Z$ plane.
We depicted the lines of constant mass, effective temperature and luminosity, as well as the instability borders for the radial mode,
$\nu_1$ (thick solid line) and for the dipole mode, $\nu_5$ (thick dashed line). We indicated masses from 6.8 up to 9.8$M_{\odot}$
and the values of $\log{T_{\rm{eff}}}$ and $\log{L/L_{\odot}}$ corresponding to the center and edges of the total error box.
Models that are located below the $\eta=0$ lines are unstable. Almost all models between the lines $\log{L/L_{\odot}}=3.654$
and $\log{L/L_{\odot}}=3.833$ are inside the observational error box of $\gamma$ Peg.

As we can see, for a given mass, models with higher metallicity require smaller values of the overshooting parameter,
effective temperature and luminosity. The same results were obtained for $\theta$ Oph (Daszy\'nska-Daszkiewicz \& Walczak 2009) and 12 Lac
(Daszy\'nska-Daszkiewicz, Szewczuk \& Walczak 2013).
Lines of constant effective temperature and luminosity are nearly parallel to each other and more steep than lines of constant mass. Having such
a huge number of seismic models fitting two frequencies of $\gamma$ Peg, we determined an approximate relation between their
parameters:
\begin{equation}
\alpha_{\rm{ov}}=-21.36(33)Z-0.2613(38)M+2.623(30).
\end{equation}

The very important point of the seismic analysis is mode instability. As can be seen in Fig.\,\ref{Z-AO-OPAL}, all models with
$Z\lesssim0.01$ have the modes $\ell=0$, p$_1$ and $\ell=1$, g$_1$ stable. It is caused by a small abundance of heavy elements,
which are crucial for exciting pulsations in the B-type stars. A lot of unstable models appear for higher value of metallicity,
$Z\gtrsim0.01$, and overshooting less than about $0.35$ for the radial mode $\nu_1$ and about $0.48$ for the dipole mode $\nu_5$.
Models with high overshooting are stable because they have too low masses and fall outside the instability region of the $\beta$ Cep
stars \citep{P99}. For more thorough discussion of the mode instability analysis see \citet{ZPDW13}.

\begin{figure*}
 \begin{center}
 \includegraphics[clip,width=160mm,height=115mm]{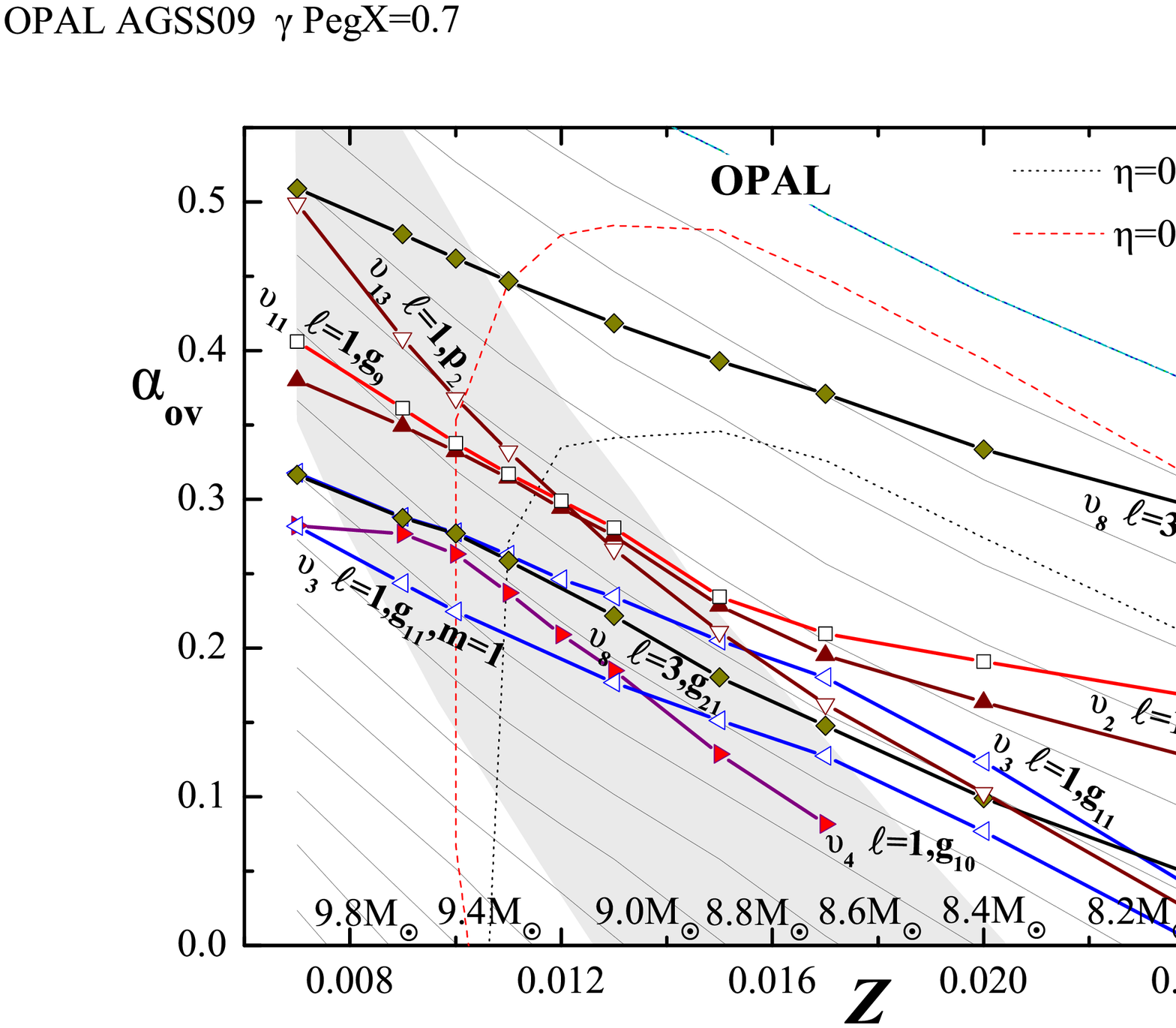}
 \caption{The same as in Fig.\,\ref{Z-AO-OPAL}, but we added lines of seismic models fitting the high-order g-mode frequencies
 which were identified  as $\ell=1$ or $\ell=3$ modes. These are: $\nu_2$ $(\blacktriangle)$, $\nu_3$ $(\lhd)$,
 $\nu_4$ $(\blacktriangleright)$, $\nu_8$ $(\blacklozenge)$ and $\nu_{11}$ $(\Box)$. The line of models fitting the p-mode frequency $\nu_{13}$ ($\ell=1$, p$_2$)
 is plotted as $(\bigtriangledown)$. The grey area indicates models located inside the total error box of $\gamma$ Peg. In the case of $\nu_3$ two values of $m$ were considered: $m=0$ and $m=1$.}
\label{Z-AO-OPAL-g}
\end{center}
\end{figure*}


\begin{figure*}
 \begin{center}
\includegraphics[clip,width=160mm,height=115mm]{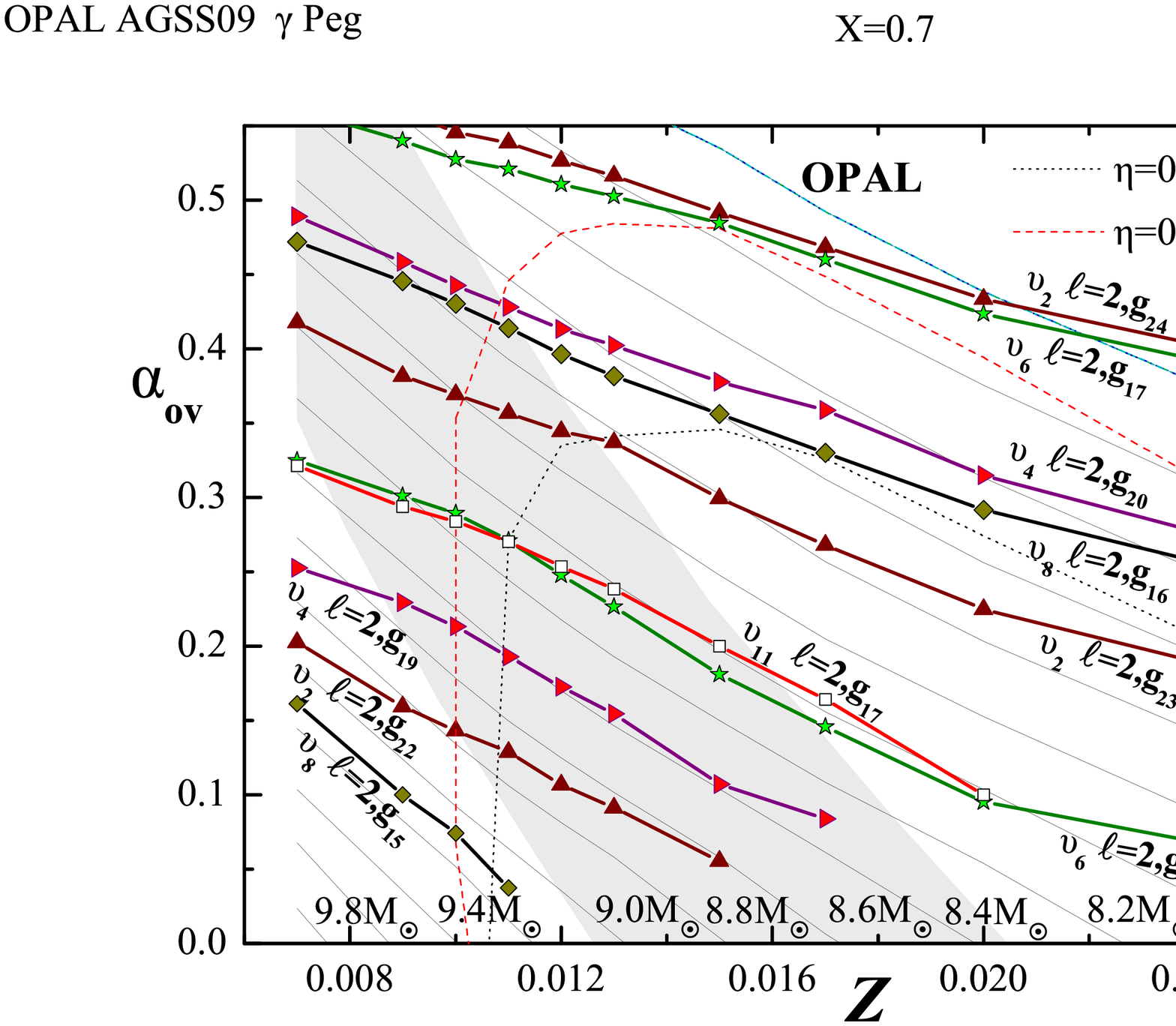}
 \caption{The same as in Fig.\,\ref{Z-AO-OPAL-g}, but models fitting frequencies identified as high-order g-modes with $\ell=2$
 were added. These are: $\nu_2$ $(\blacktriangle)$, $\nu_4$ $(\blacktriangleright)$, $\nu_6$ $(\bigstar)$, $\nu_8$ $(\blacklozenge)$
 and $\nu_{11}$ $(\Box)$.}
\label{Z-AO-OPAL-g-2}
\end{center}
\end{figure*}

In Fig.\,\ref{Z-AO-OPAL-g} and \ref{Z-AO-OPAL-g-2}, we plotted the same as in Fig.\,\ref{Z-AO-OPAL}, but we marked also lines
of models fitting some of seven frequencies: $\nu_2$, $\nu_3$, $\nu_4$, $\nu_6$, $\nu_8$, $\nu_{11}$ and $\nu_{13}$.
In most cases, we assumed that these modes are axisymmetric ($m=0$) because there is no determination of $m$. For $\nu_3$ we considered two possibilities, $m=0$ and $m=1$, since there are some indications that the mode is prograde \citep{Pandey_al2011}. We considered simple rotational splitting in order to calculate the centroid frequency, that is $\nu_{n\ell m}=\nu_{n\ell}+m(1-C_{n\ell})\nu_{\rm{rot}}$, where $C_{n\ell}$ is the Ledoux constant and $\nu_{\rm{rot}}$ is the rotational frequency.

Models located on each line fit simultaneously three frequencies.
Fig.\,\ref{Z-AO-OPAL-g} includes frequencies which were identified as $\ell=1$ or 3, whereas Fig.\,\ref{Z-AO-OPAL-g-2} - those
which have the degree $\ell=2$. These seven frequencies were fitted with an accuracy of $10^{-4}$ - $10^{-3}$ c/d, equivalent to the observational errors.

In the case of high-order g-modes ($\nu_2$, $\nu_3$, $\nu_4$, $\nu_6$, $\nu_{8}$, $\nu_{11}$), we usually had to include
more than one radial order. In a considered range of metallicity and overshooting parameters, the frequency $\nu_3$ is
the $\ell=1$, g$_{11}$ mode (in both cases, for $m=0$ and $m=1$) and the frequency $\nu_6$ is the $\ell=2$, g$_{16}$ or g$_{17}$ mode. Identification of the mode degree
of $\nu_2$, $\nu_4$, $\nu_8$ and $\nu_{11}$ is ambiguous and we had to consider more possibilities.
For $\nu_2$ we got the modes $\ell=1$, g$_{12}$ and $\ell=2$, g$_{22}$, g$_{23}$, g$_{24}$,
for $\nu_{11}$ - the modes $\ell=1$, g$_9$ and $\ell=2$, g$_{17}$. The frequency $\nu_8$ can be a mode with $\ell=2$ or $\ell=3$.
In the former case, it has to be the g$_{15}$ or g$_{16}$ mode and in the later - the g$_{21}$ or g$_{22}$ mode. If the frequency
$\nu_4$ is a dipole, then it has to be g$_{10}$, but if it is a quadruple, it is g$_{19}$ or g$_{20}$.
Because in many models the pressure mode $\ell=1$, p$_2$ has a frequency close to $\nu_{13}$, we plotted also a model line
reproducing this frequency.

Models fitting three frequencies require higher overshooting, effective temperature, luminosity and mass if metallicity decrease.
Unfortunately, high-order g-modes are usually stable. The modes $\nu_{2}$($\ell=2$, g$_{23}$) and $\nu_{6}$($\ell=2$, g$_{16}$)
are exited if the metallicity parameter $Z\gtrsim0.02$; $\nu_{2}$($\ell=2$, g$_{24}$), $\nu_{4}$($\ell=2$, g$_{20}$) and
$\nu_{8}$($\ell=3$, g$_{21}$) if $Z\gtrsim0.016$ and $\nu_{6}$($\ell=2$, g$_{17}$) and $\nu_8$($\ell=3$, g$_{22}$) if $Z\gtrsim0.013$.

Moreover, we were able to find a lot of models fitting four frequencies and two models fitting five frequencies.
A model with $M=8.093M_{\odot}$, $\log{T_{\rm{eff}}}=4.3286$, $\alpha_{\rm{ov}}=0.27$ and $Z=0.011$ fits the modes
$\nu_{1}$($\ell=0$, p$_{1}$), $\nu_{5}$($\ell=1$, g$_1$), $\nu_{3}$ ($\ell=1$, g$_{11}$, $m=0$), $\nu_{6}$ ($\ell=2$, g$_{16}$)
and $\nu_{11}$ ($\ell=2$, g$_{17}$). We will call it MODEL1 and it is marked with a diamond in
Fig.\,\ref{HR}. The parameters of the other model fitting five modes are $M=7.895M_{\odot}$,
$\log{T_{\rm{eff}}}=4.31956$, $\alpha_{\rm{ov}}=0.30$ and $Z=0.012$. The fitted modes are: $\nu_{1}$($\ell=0$, p$_{1}$), $\nu_{5}$($\ell=1$, g$_1$), $\nu_{2}$
($\ell=1$, g$_{12}$), $\nu_{11}$ ($\ell=1$, g$_{9}$) and $\nu_{13}$ ($\ell=1$, p$_{2}$).
It is worth to note that these two models are located inside the observational error box of $\gamma$ Peg.
However, neither of these two models can reproduce the frequency $\nu_{4}$, what can indicate
that it is not an axisymmetric mode, as we have assumed.

\begin{figure*}
\begin{center}
 \includegraphics[clip,width=160mm,height=75mm]{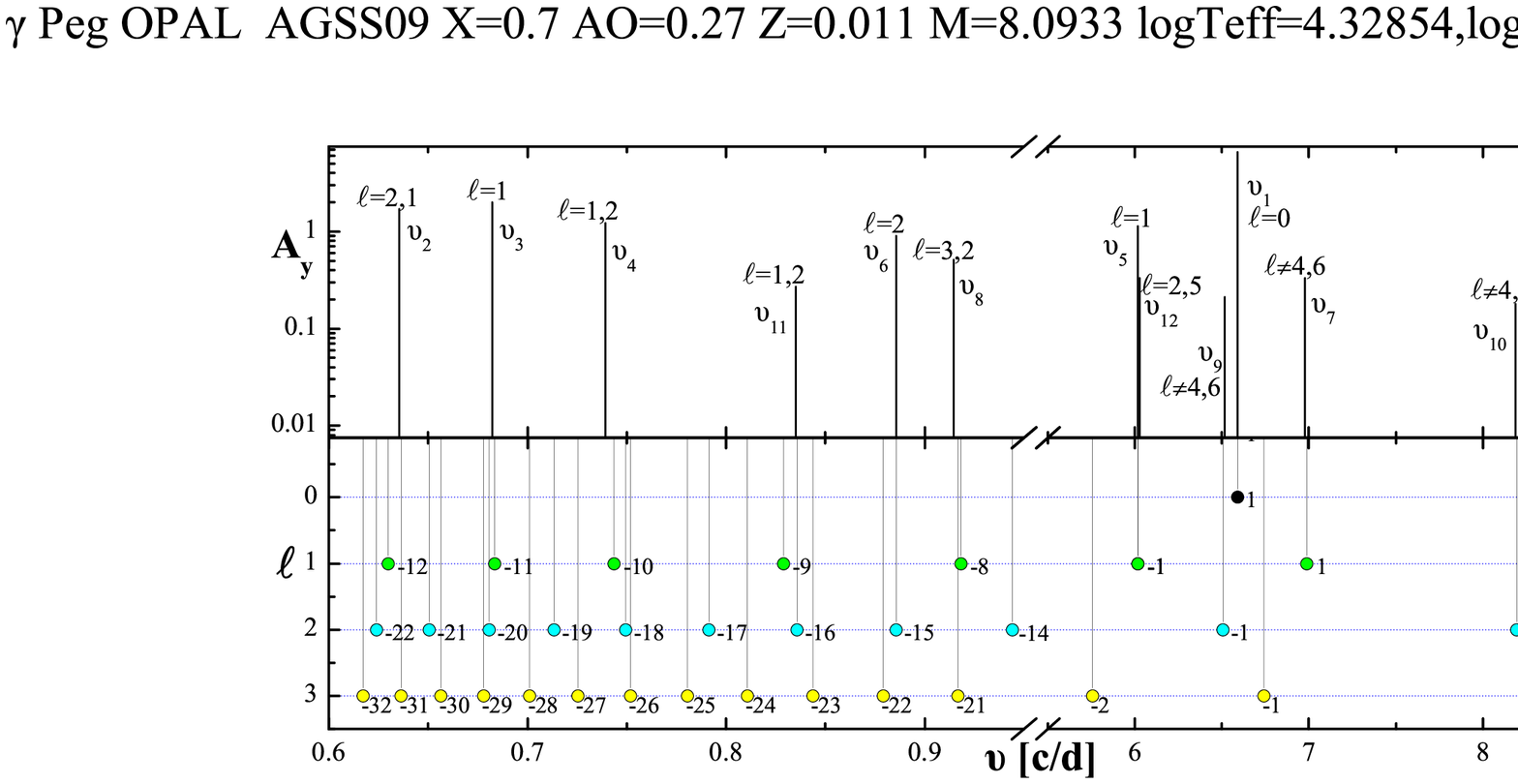}
 \caption{Comparison of the observed frequencies of $\gamma$ Peg (upper panel) with the theoretical counterparts (bottom panel)
corresponding to MODEL1 of $\gamma$ Peg which has the following parameters: $M=8.093M_{\odot}$, $\log{T_{\rm{eff}}}=4.3286$,
$\alpha_{\rm{ov}}=0.27$ and $Z=0.011$. The height of lines in the upper panel corresponds to the amplitude in the Str\"omgren y passband
while in the bottom panel to the mode degree, $\ell$. The radial order is given at each theoretical frequency peak.
The values of $\ell$ shown in the upper panel were derived in Section\,\ref{Ident}.}
\label{frespec}
\end{center}
\end{figure*}

In Fig.\,\ref{frespec}, we present the detailed comparison of the observational frequency spectrum and their possible theoretical
counterparts for MODEL1 described above. We considered modes with $\ell$ from 0 to 3. Each theoretical frequency peak is labeled with the radial order, $n$.
The well identified frequencies have the best theoretical counterparts
for mode degrees generally consistent with the photometric identification. The exception is $\nu_{2}$. Despite the fact that it was
identified as a dipole or quadruple mode, the best match is for the $\ell=3$ mode. A comparison with the theoretical spectrum may suggest,
that this is the prograde mode ($m>0$).
The frequency $\nu_{12}$ can be the prograde mode, $\ell=1$, m=-1, g$_1$ as assumed by Handler et al.\,(2009). The frequencies $\nu_9$ and $\nu_{10}$ seems to be
the $\ell=2$, g$_1$ and p$_0$ modes, respectively. The frequencies $\nu_7$ and $\nu_{13}$ can be consecutive dipole modes, i.e., p$_1$ and p$_2$.
We can see also, that the frequency $\nu_{14}$ is consistent with the first overtone radial mode, but, as we mentioned in Section\,\ref{Ident}, it does not agree with our identification. The other close mode to $\nu_{14}$ is $\ell=6$, g$_1$ (not shown in Fig.\,6), which is not inconsistent with photometric identification.

\begin{figure*}
\begin{center}
 \includegraphics[clip,width=83mm,height=70mm]{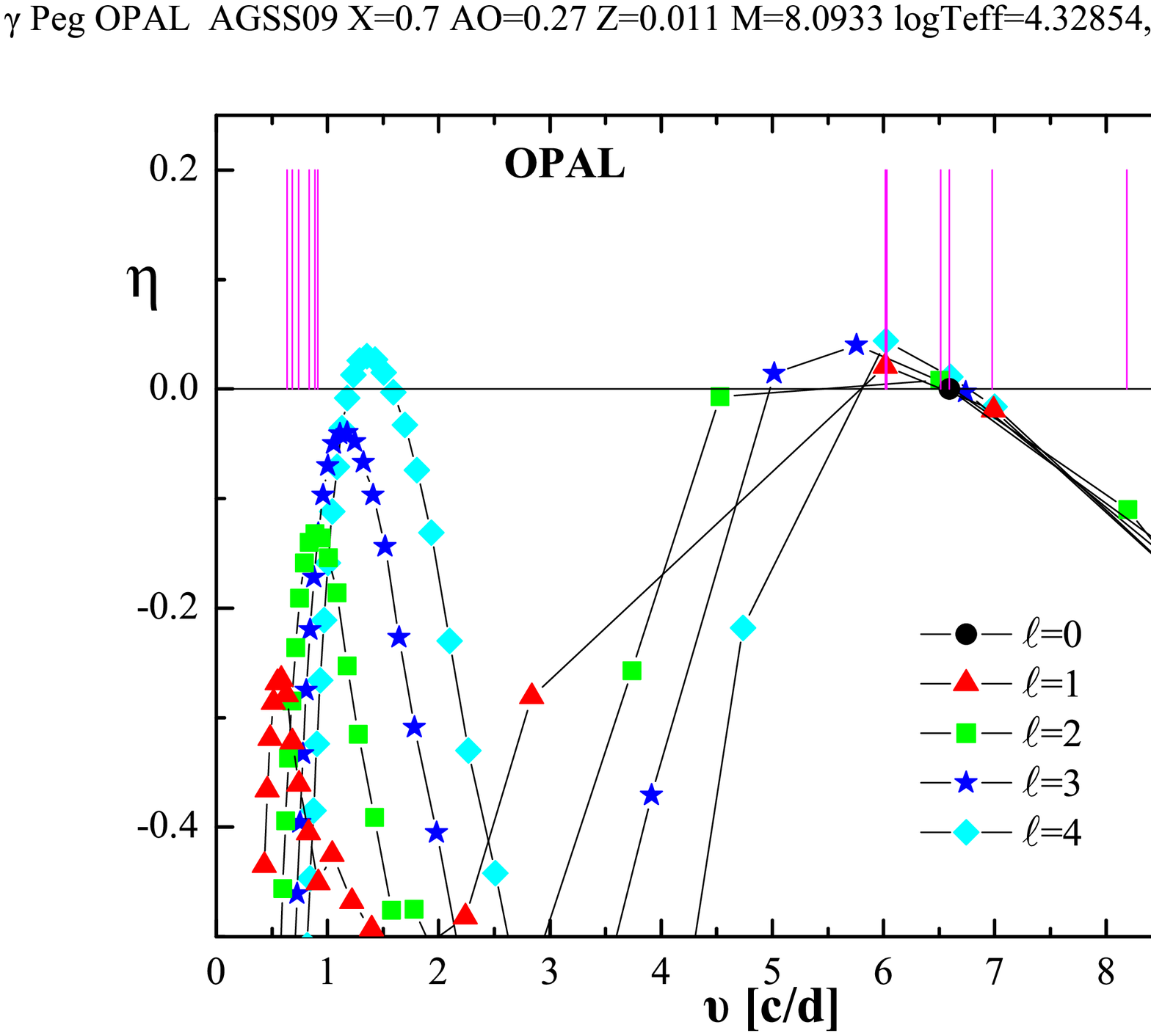}
 \includegraphics[clip,width=83mm,height=70mm]{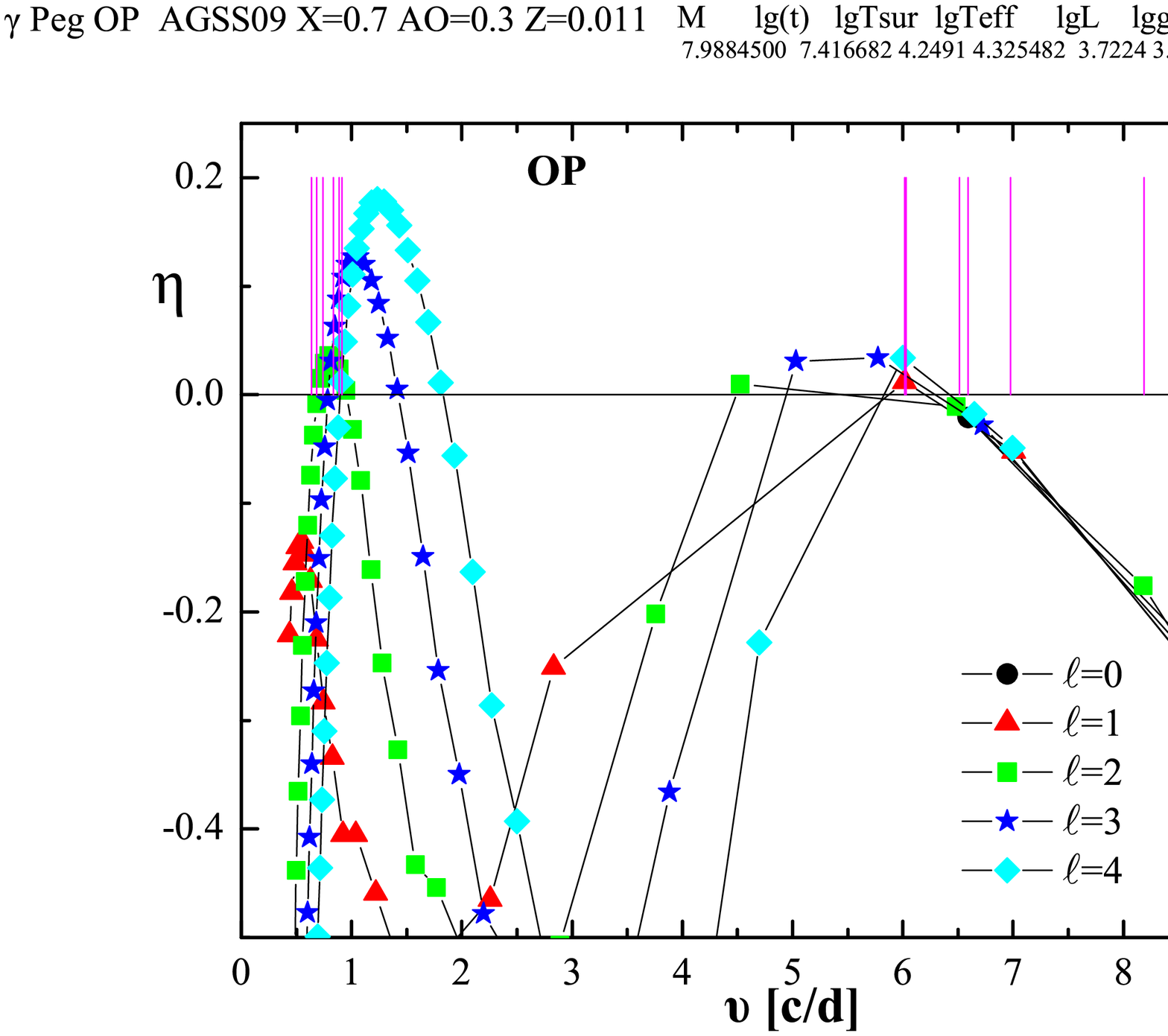}
 \caption{Run of the instability parameter, $\eta$, with frequency for MODEL1 (the left panel) and a model with very similar parameters
  but computed with the OP tables (the right panel). The vertical lines corresponds to the observed oscillation spectrum of $\gamma$ Peg.}
\label{etafreq}
\end{center}
\end{figure*}

Let us now check the instability conditions for MODEL1. In the left panel of Fig.\,\ref{etafreq}, we plotted the instability
parameter, $\eta$, as a function of frequency for modes with $\ell=0-4$. For a comparison, in the right panel of Fig.\,\ref{etafreq},
we plotted the same for a model with very similar parameters but computed with the OP data. Short, vertical lines represent
the observed frequencies of $\gamma$ Peg. Modes with $\eta>0$ are excited in the model. As we can see, with the OPAL opacities
only frequencies in the range from about 5 up to 7 c/d are unstable. In the domain of the high order g-modes only high-degree modes
($\ell\ge 4$) are excited with frequencies $\nu\gtrsim1$ c/d. The frequencies in the range 8-9.5 c/d and 0.6-0.9 c/d are stable.
Slightly better situation appears with the OP data. The model computed with these opacity data excite high-order g-modes with
$\ell\ge 2$. But neither with the OP nor OPAL date we were able to excite modes with frequencies higher than about 7 c/d.
It is tempting to explain this unsatisfactory results assuming higher opacities in the driving zones, but, as we will discuss
it in the Section\,\ref{opac_enh}, it causes other problems.

\subsection{Fitting the nonadiabatic parameter, $f$}

As we have mentioned in the Introduction, an adequate seismic model should reproduce not only pulsational
frequencies but also other characteristic parameters. Such parameter, which is determinable from observations,
is the $f$-parameter associated with each frequency. The theoretical values of $f$ depend mostly on properties of the driving zone.

To derive the empirical values of the $f$-parameter, we used the LTE models of stellar atmospheres
with the microturbulent velocity of $\xi_t=2$ km/s. The results do not differ qualitatively from those computed with the non-LTE
atmosphere models \citep{Lanz2007} as well as with higher microturbulent velocity $\xi_t=8$ km/s.

\begin{figure*}[h]
\begin{center}
 \includegraphics[clip,width=83mm,height=70mm]{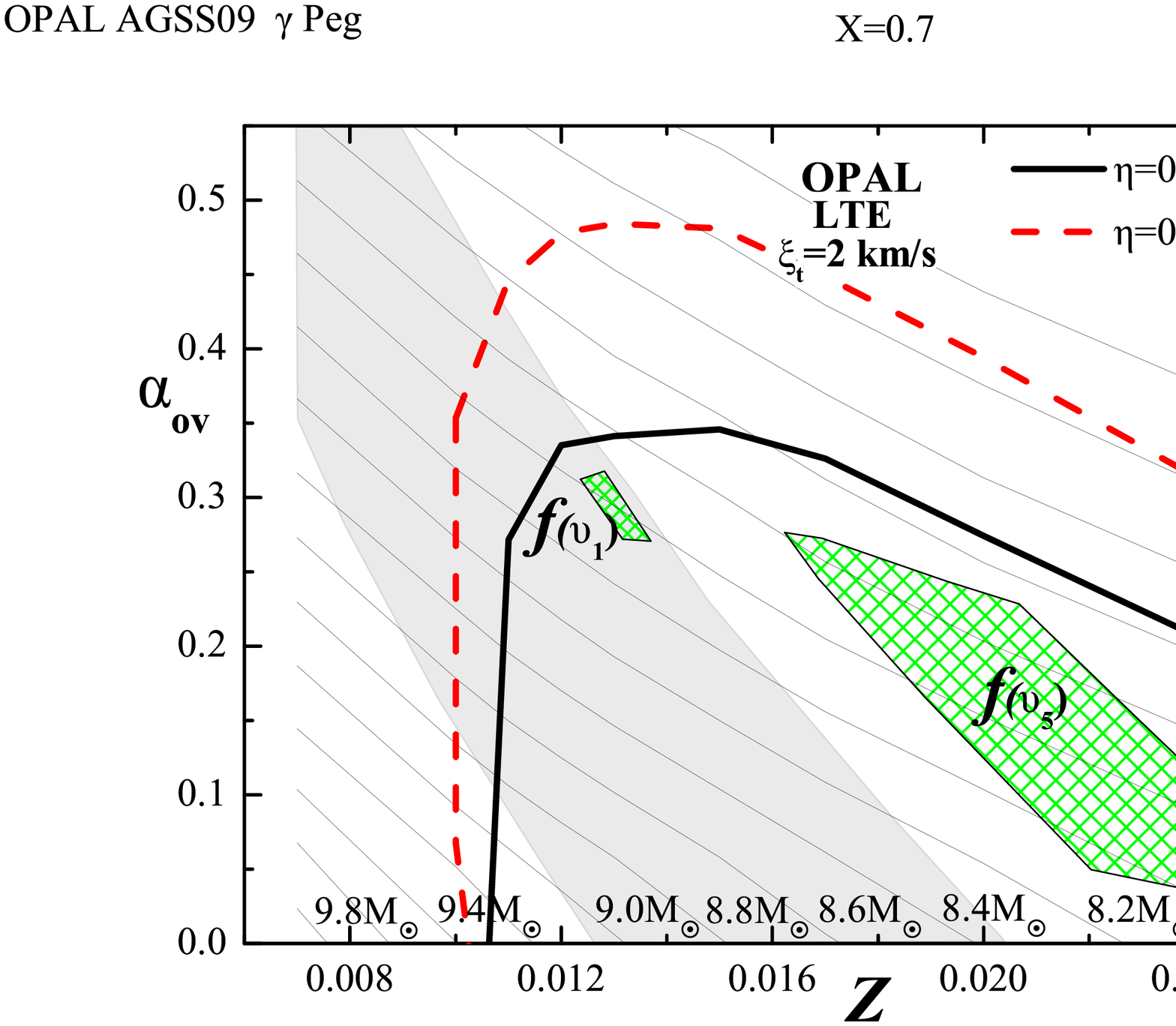}
 \includegraphics[clip,width=83mm,height=70mm]{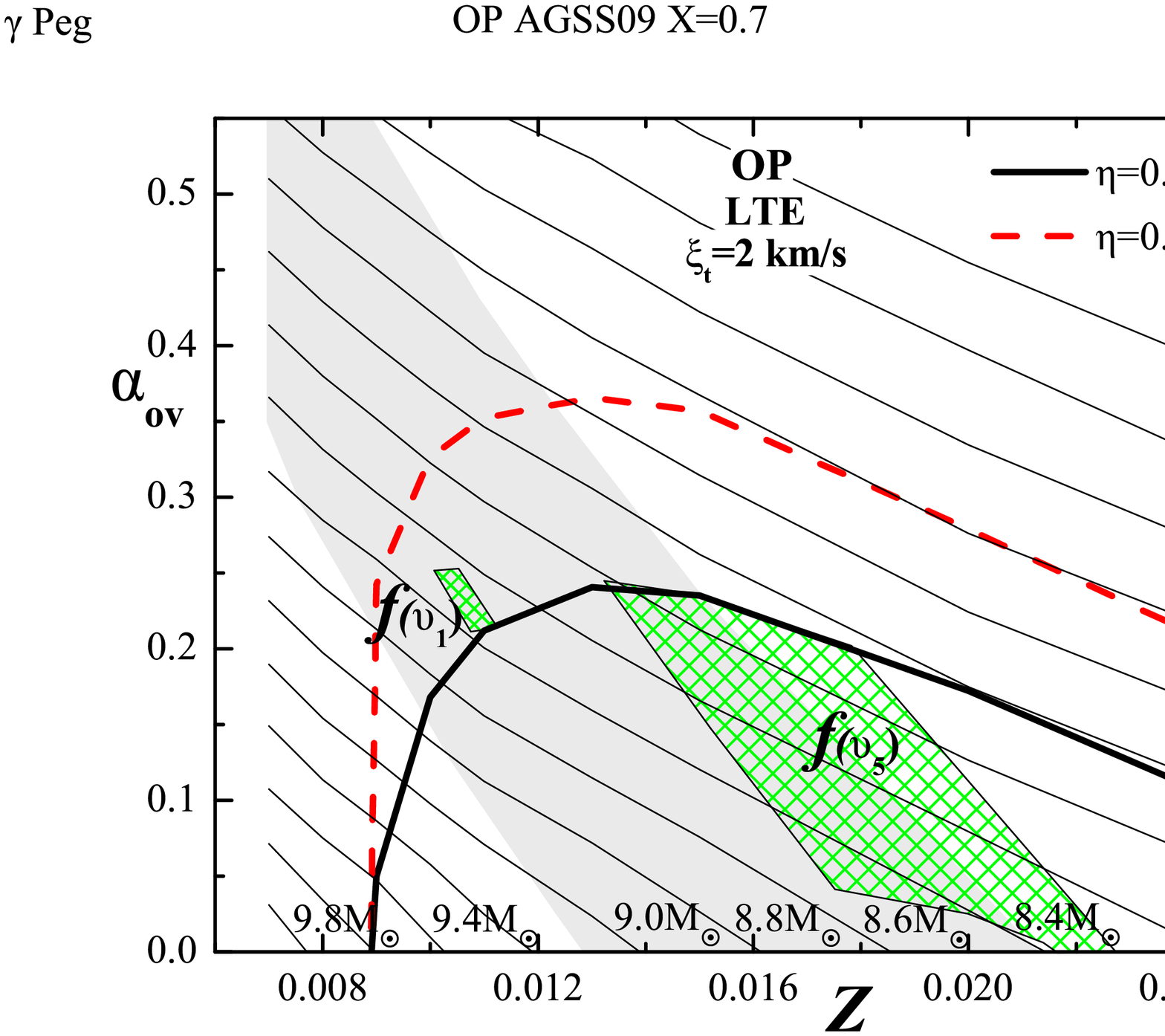}
 \caption{The same as in  Fig.\,\ref{Z-AO-OPAL} but we marked models fitting the empirical values of the nonadiabatic $f$-parameter (hatched areas) of the radial fundamental mode (labeled as $f(\nu_1$)) and of the dipole mode g$_1$ (labeled as $f(\nu_5$)). In the left panel we used the OPAL data and in the right panel the OP data.}
\label{Z-AO-OPAL-OP}
\end{center}
\end{figure*}

In Fig.\,\ref{Z-AO-OPAL-OP}, we put seismic models fitting, within the observational errors, the empirical values of the $f$-parameter
(hatched areas) for the radial fundamental mode (labeled as $f(\nu_1$)) and for the dipole g$_1$ mode (labeled as $f(\nu_5$)).
The area $f(\nu_5)$ is larger than $f(\nu_1)$ because of larger observational errors in the photometric amplitudes and phases
of the frequency $\nu_5$. The left and right panels of Fig.\,\ref{Z-AO-OPAL-OP} correspond to computations with the OPAL and OP tables,
respectively. As we can see, the requirement of fitting the $f$-parameter for the $\nu_1$ mode reduce significantly the allowed range
of stellar parameters of $\gamma$ Peg. With the OPAL tables and the hydrogen abundance $X=0.7$, only models with
$\alpha_{\rm{ov}}\approx0.3$ and $Z\approx0.013$ can be considered.
The OPAL models fitting the $f$-parameter for the $\nu_1$ mode are inside the observational error box of $\gamma$ Peg,
while all models fitting the $f$-parameter for the $\nu_5$ mode are outside the box. In the case of the OP models, the areas $f(\nu_1)$ and part of $f(\nu_5)$
are inside the error box.

Results for the OPAL and OP tables are similar and the biggest difference is in the position of the instability borders.
For the OP models, these lines are shifted to the lower values of metallicity, $Z$, and the overshooting parameter, $\alpha_{\rm{ov}}$.
Furthermore, for a given value of $Z$ and $\alpha_{\rm ov}$, models with the OP tables have slightly higher mass whereas effective temperature
and luminosity are almost the same. Moreover, the areas indicating models fitting the $f$-parameter appear at the lower overshooting and metallicity.
Unfortunately, there is no model fitting the $f$-parameter for
both $\nu_1$ and $\nu_5$, simultaneously; neither with the OPAL nor OP data.

In the case of the most high-order g-modes, the empirical and theoretical values of the real part of $f$ agreed and the imaginary part differed significantly.
This occurs both with the OPAL and OP opacities. Only for a mode $\nu_6$ we were able to find some seismic models which fit its empirical value of $f$
(both with the OPAL and OP data). These models, obtained with the OP data, have $\alpha_{\rm ov}\approx 0.31$ and $Z\approx 0.008-0.009$, and
are located inside the observational error box of $\gamma$ Peg. However, they do not overlap neither with the area $f(\nu_1)$ nor $f(\nu_5)$.
The OPAL models reproducing the $f$-parameter of $\nu_6$ are outside the allowed range of parameters and have $\alpha_{\rm ov}\approx 0.51$ and $Z\approx 0.010-0.012$.

In the next step, we checked the effects of the hydrogen abundance and the heavy elements composition. In comparison with the results computed with
the standard hydrogen abundance, $X=0.7$, models with $X=0.75$ for a given value of $Z$ and $\alpha_{\rm{ov}}$ have larger masses
(about $0.3M_{\odot}$), smaller effective temperatures ($\Delta\log{T_{\rm{eff}}}\sim0.18$) and smaller luminosities
($\Delta\log{L/L_{\odot}}\sim0.04$). Also models fitting the $f$-parameters required much smaller value of the overshooting parameter.
Still, there were no models fitting the $f$-parameter for both $\nu_{1}$ and $\nu_5$ frequency simultaneously.

\begin{figure}[h]
\begin{center}
 \includegraphics[clip,width=83mm,height=70mm]{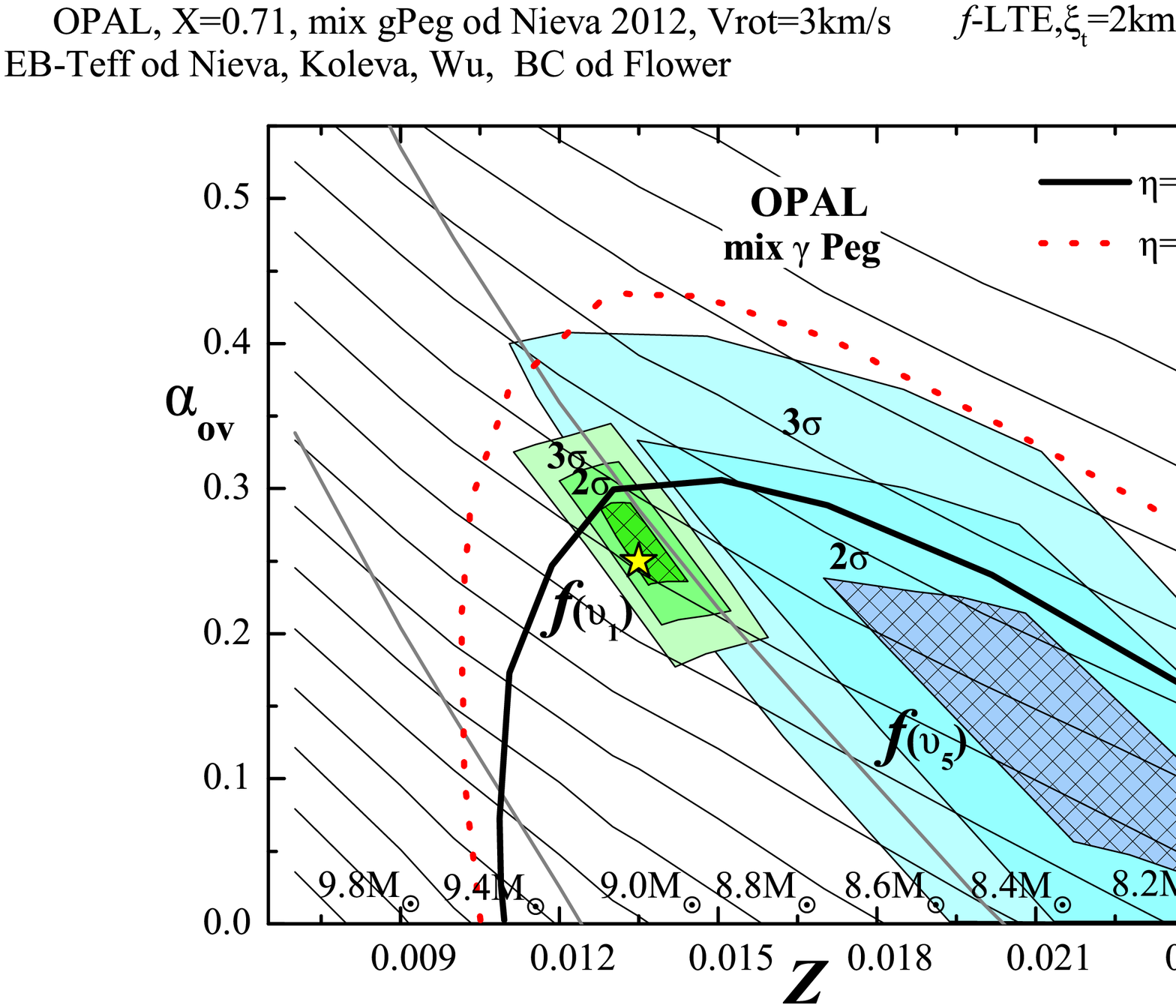}
 \caption{The same as in the left panel of Fig.\,\ref{Z-AO-OPAL-OP} but for the chemical composition of $\gamma$ Peg as determined by \citet{Nieva+2012}. The dark shaded-in and bright shaded-in areas around the hatched regions indicate models fitting the $f$-parameters within 2$\sigma$ and 3$\sigma$, respectively.
 The asterisk marks model chosen for a comparison of the values of $f$.}
\label{Z-AO-OPAL-mixgPeg}
\end{center}
\end{figure}

Then, we changed the chemical mixture. In Fig.\,\ref{Z-AO-OPAL-mixgPeg}, we show seismic models calculated with the chemical composition of $\gamma $ Peg as determined by \cite{Nieva+2012}. The $\gamma$ Peg mixture has smaller abundance of iron-group elements than the Sun. These elements are extremely important for exciting pulsation and because of their deficiency, the instability region is smaller. Nevertheless, there are still a lot of unstable models fitting two pulsational frequencies and the $f$-parameter corresponding to $\nu_1$ or $\nu_5$. In this case seismic models fitting the values of the $f$-parameter require slightly less efficient core overshooting.

Because the formal errors of $f$ can be underestimated, in Fig.\,\ref{Z-AO-OPAL-mixgPeg}, we marked additionally models fitting the $f$-parameters within 2$\sigma$, (shaded-in, dark areas around the hatched regions) and 3$\sigma$ (shaded-in, bright areas around the hatched regions), where $\sigma$ is the empirical error of $f$. Now, the majority of models reproducing $f$ for $\nu_1$ fit also, within 3$\sigma$, the $f$-parameter corresponding to $\nu_5$.

\begin{figure*}
\begin{center}
 \includegraphics[clip,width=83mm,height=70mm]{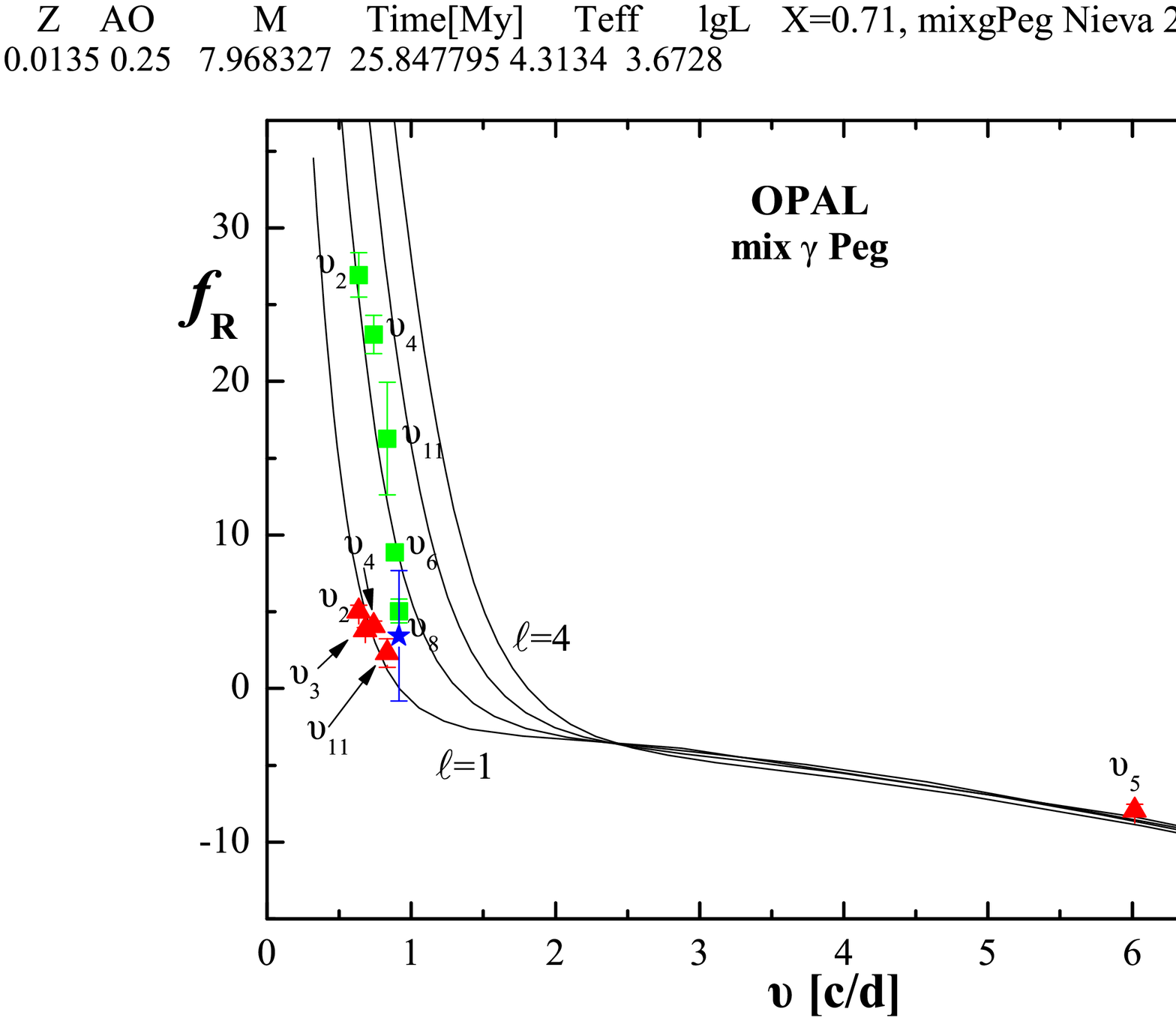}
 \includegraphics[clip,width=83mm,height=70mm]{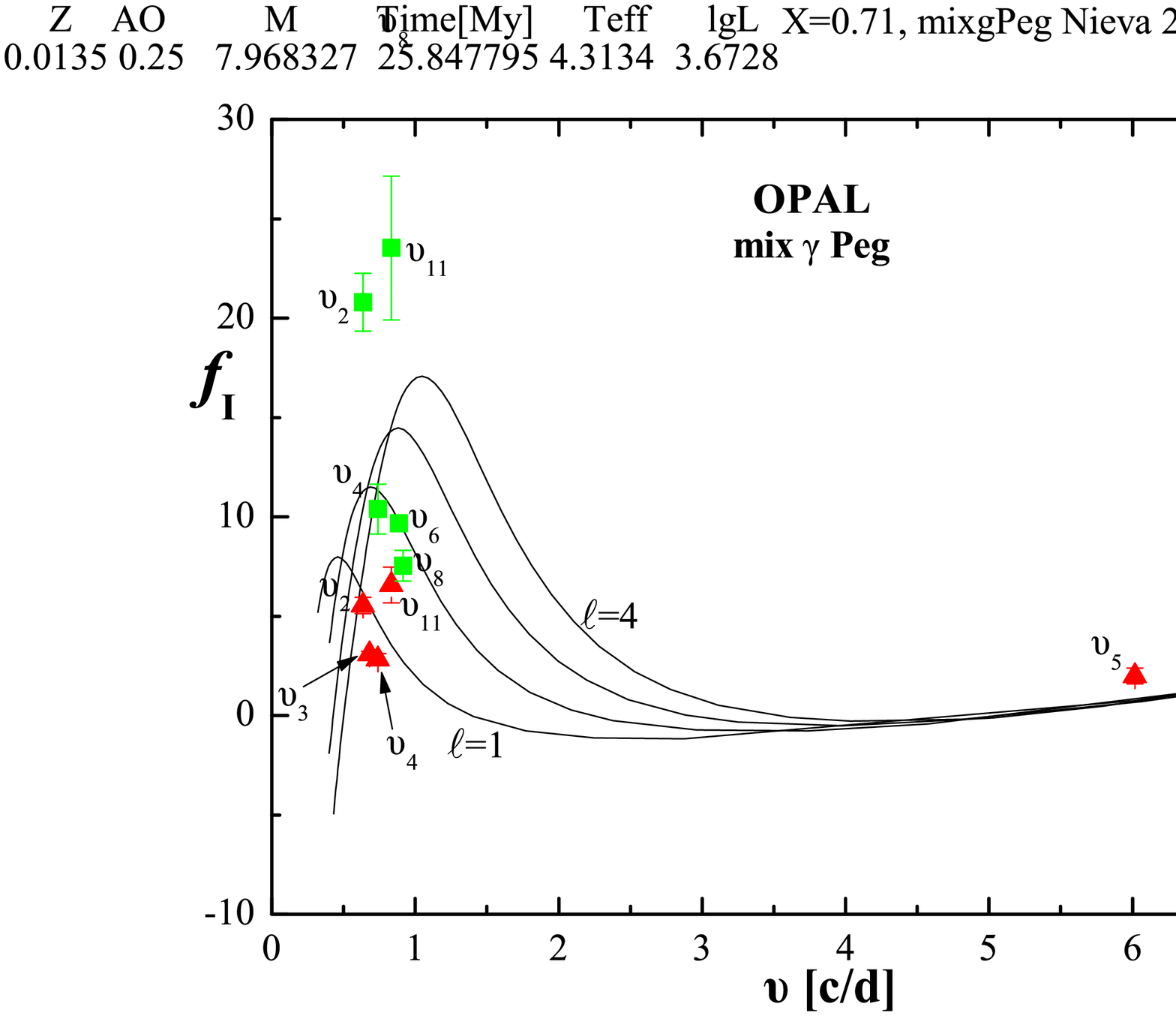}
 \caption{The real (left panel) and imaginary (right panel) part of the nonadiabatic parameter $f$ as a function of pulsational frequency for MODEL2
  of $\gamma$ Peg marked as asterisk in Fig.\,1 and Fig.\,9. The empirical values of $f$ were put for the eight well identified frequencies; circles correspond to $\ell=0$, triangles to $\ell=1$, squares to $\ell=2$ and stars to $\ell=3$.}
\label{fr-fi-ni-OPAL-mixgPeg}
\end{center}
\end{figure*}

It is important to add, that in all considered cases, we were able to find also a model fitting, within 1$\sigma$ error,
the $f$- parameter for $\nu_1$ and the real part of $f$ for $\nu_5$. The disagreement remains for the imaginary part of $f$.
This is clearly visible in Fig.\,\ref{fr-fi-ni-OPAL-mixgPeg}, where we compare the theoretical and empirical values of $f$ in a function of the frequency.
The left and right panels show the real ($f_{\rm{R}}$) and imaginary ($f_{\rm{I}}$) part, respectively.
The parameters of the chosen seismic model marked with an asterisk in Fig.\,\ref{HR} and \ref{Z-AO-OPAL-mixgPeg} are:
$M=7.968M_{\odot}$, $\log{T_{\rm{eff}}}=4.3134$, $\log{L/L_{\odot}}=3.6728$, $X=0.71$, metallicity $Z=0.0135$,  $\alpha_{\rm{ov}}=0.25$,
the OPAL data and $\gamma$ Peg mixture. We called it MODEL2. Modes with the degrees $\ell=0,1,2,3,4$ were considered.

We marked the empirical values of the $f$-parameters for $\nu_1$, $\nu_{5}$ and six high-order g-modes: $\nu_2$ ($\ell=2$ and 1),
$\nu_3$ ($\ell=1$), $\nu_4$ ($\ell=2$ and 1), $\nu_6$ ($\ell=2$), $\nu_8$ ($\ell=2$ and 3), $\nu_{11}$ ($\ell=1$ and 2).
The value of the imaginary part of the empirical $f$-parameter for $\nu_8$ ($\ell=3$) is out of the scale.
The empirical values of $f$ corresponding to different degrees have different symbols.
We can conclude that only in the case of the real part of $f$ an agreement between the theoretical and empirical values of $f$
is quite good. The same result was obtained by \citet{DDDP05} and \citet{DW10} for $\nu$ Eri.

The values of the $f$-parameter for p modes are almost independent of the mode degree, $\ell$, and change slowly with the mode frequency,
while for high-order g-modes $f$-parameter depend strongly on $\ell$ and changes rapidly with $\nu$. The fact, that the empirical values
of the real part of $f$, especially for high-order g-modes, are located almost perfectly along the lines of theoretical counterparts
are very encouraging for further studies and it shows a great potential of the $f$-parameter, particularly for the SPB-type modes.

The empirical values of the nonadiabatic $f$-parameter, plotted in Fig.\,\ref{fr-fi-ni-OPAL-mixgPeg}, are listed in Table\,\ref{frfi}.
In the fifth column we give also the values of the intrinsic amplitude, $|\varepsilon|$, multiplied by spherical harmonic, $Y_{\ell}^{m}(i,0)$, where $i$ is the inclination angle. In the case of the radial mode, $\nu_1$, we have the exact value of $|\varepsilon|$, whereas for other modes we can get only their lower limit. As we can see, the intrinsic amplitude of the dominant radial mode is very small and amounts to $\sim$0.26 per cent of the stellar radius and it is about three times larger than for other modes.
The last column contains the values of the discriminant, $\chi^2_E$, which measures goodness of the fit between the theoretical
and observational values of the photometric amplitudes and phases.
\begin{table}
\begin{center}
\caption{The empirical values of  the real and imaginary part of $f$,  and the absolute value of the intrinsic mode amplitude $|\tilde{\varepsilon}|$ for 8 frequencies of $\gamma$ Peg. Columns from left to right are: pulsational frequency, mode degree, $\ell$, empirical values of the real, $f_{\rm{R}}$, and imaginary, $f_{\rm{I}}$, part of the nonadiabatic parameter, the intrinsic amplitude multiplied by spherical harmonic, $|\tilde{\varepsilon}|=|\varepsilon Y_{\ell}^{m}(i,0)|$, and the discriminator, $\chi^2$, describing the goodness of the fit. These quantities were derived for the parameters of MODEL2, marked with asterisk in Fig.\,\ref{HR} and \ref{Z-AO-OPAL-mixgPeg}.}
\begin{tabular}{llrrllll}
\hline
frequency &\multirow{2}{*}{$\ell$}&\multirow{2}{*}{$f_{\rm{R}}$}&\multirow{2}{*}{$f_{\rm{I}}$}& \multirow{2}{*}{$|\tilde{\varepsilon}|=|\varepsilon|Y_{\ell}^{m}(i,0)$}&\multirow{2}{*}{$\chi^2_E$}\\
$~~~~[$c/d$]$&& &&                               \\
\hline
$\nu_1=$6.58974&0&-9.43(11)&1.46(11)&0.002697(12)&4.50 \\\hline
\multirow{2}{*}{$\nu_2=$0.63551}&1&5.03(42)&5.54(41)&0.000584(11)&1.70\\
&2&26.9(1.5)&20.8(1.5)&0.000376(6)&1.40 \\\hline
$\nu_3=$ 0.68241&1&3.81(15)&3.08(15)&0.000910(7)&0.54 \\\hline
\multirow{2}{*}{$\nu_4=$0.7394}&1&4.10(29)&2.84(29)&0.000684(10)&1.15 \\
&2&23.1(1.3)&10.4(1.3)&0.000451(8)&1.52 \\\hline
$\nu_5=$ 6.01616&1&-7.94(41)& 1.99(40)&0.000381(7)&0.62 \\\hline
$\nu_6=$0.8855&2&8.87(37)& 9.67(37)&0.000734(4)&0.37 \\\hline
\multirow{2}{*}{$\nu_8=$0.91442}&2&5.05(78)& 7.54(77)&0.000485(6)&0.74 \\
&3&3.4(4.3)&56.7(4.2)&0.000717(7)&0.41 \\\hline
\multirow{2}{*}{$\nu_{11}=$0.8352}&1&2.31(92)&6.57(90)& 0.000255(12)&1.50\\
 &2&16.3(3.7)&23.5(3.6)& 0.000172(9)&1.77 \\
\hline
\end{tabular}
\label{frfi}
\end{center}
\end{table}

\subsection{Effect of the opacity enhancement}
\label{opac_enh}

In the previous section, we have tried to find seismic models of $\gamma$ Peg reproducing two frequencies and corresponding values of the $f$-parameter.
Using the standard opacity data and 1$\sigma$ error in $f$ we did not succeed in finding a model fitting the values of $f$ for $\nu_1$ and $\nu_5$ simultaneously. Because of this, we decided to change artificially the OPAL opacity tables. Two cases were examined.

In the first case we tested the 50\% opacity enhancement near the $Z$-bump, i.e., at $\log{T}\approx5.3-5.5$.
The lines of constant effective temperature, luminosity and mass on the $\alpha_{\rm{ov}}~ vs.~ Z$ plane were nearly unchanged in
comparison with the standard models, but opacity enhancement had a huge impact on the instability borders. Almost all our models had
the modes $\ell=0$, p$_{1}$ and $\ell=1$, g$_1$ unstable for metallicity as low as $Z=0.007$. Unfortunately, with the modified
opacities no model reproduces the empirical values of $f$, neither for $\nu_1$ nor for $\nu_5$.

As we have already mentioned, the theoretical values of the real part of $f$ calculated with the standard opacities (OP, OPAL) agreed very well
with their empirical counterparts for almost all modes. The problem was to fit the imaginary part of $f$.
The change of the opacity near the $Z$-bump significantly altered the real part of $f$, whereas the imaginary part was almost unaffected.

In the second case, opacities near the Deep Opacity Bump (DOB), occurring in the temperature range of $\log{T}=6.2-6.5$, were increased by 20\%. This modification did not change our models significantly. Also the instability regions for the modes $\nu_1$ and $\nu_5$
were nearly unaffected. This is caused by the fact that this bump is located deep inside the star and contains a small amount of mass. Therefore, this opacity modification did not substantially change the star structure. The only noticeable effect was the small reduction of the overshooting parameter of models fitting the $f$-parameters for both $\nu_1$ and $\nu_5$ modes. However, there is still no seismic model fitting the $f$-parameter for the modes $\nu_1$ and $\nu_5$ simultaneously.
The effect of the opacity enhancement on the frequencies and their instabilities will be discussed in details
in our next paper \citep{ZPDW13}.

\section{Conclusions}

The aim of this paper was to give a more detailed interpretation of the oscillation spectrum of the hybrid pulsator $\gamma$ Pegasi.
We began with identification of the mode degree, $\ell$, for the 14 frequencies:  8 of the $\beta$ Cep type and 6 of the SPB type.
Based on the two approaches, we were able to determine unambiguously four frequencies. For the other five frequencies two possible values of $\ell$ were obtained.
In the case of the remaining five frequencies only some constraints were derived.

Then, we tried to construct seismic models which fit the two low order p/g-mode frequencies and their corresponding values of the nonadiabatic complex parameter $f$.
We chose the frequencies $\nu_1$ and $\nu_5$, which were identified as the radial fundamental mode and dipole g$_1$ mode, respectively.
The problem we encountered was that there was no seismic model reproducing the $f$-parameter for these two frequencies simultaneously.
This inconsistence can be caused either by the underestimated errors or/and indicate that some additional effects should be included in pulsation modeling.
One of the reasons could be inadequacies in the opacity data.
\cite{ZP2009} have suggested increasing opacity by 20-50\% around the Z-bump and DOB (Deep Opacity Bump) to explain the observed frequency range of $\gamma$ Peg.
However, our studies showed that these artificially increased opacities spoiled even more the agreement between the empirical and theoretical values of the $f$-parameter.
With the modified opacities there is no model fitting the empirical values of $f$, neither for $\nu_1$ nor $\nu_5$.

Although we did not fully succeeded in constructing complex seismic models of $\gamma$ Peg, we have showed directions and problems that need to be solved.
There are the two main messages from this paper. The first one is a need for more accurate data on multi-colour time-series photometry and radial velocity data
to better identify the observed modes and determine the empirical values of the $f$-parameter. The second one is that if opacities are to ,,blame'' for these disagreements
and problems with mode instability, the improvement in computations of this microphysics data should be done in a more sophisticated way.


\section*{Acknowledgments}
We gratefully thank Gerald Handler for kindly providing data on photometric and radial velocity variations.
PW's work was supported by the Human Capital Programme grant financed by the European Social Fund.
AAP and TZ acknowledge partial financial support from the Polish NCN grants 2011/01/B/ST9/05448 and 2011/01/M/ST9/05914.

\end{document}